\documentclass[superscriptaddress,floatfix,aps,pre,twocolumn,notitlepage,shortbibliography,nofootinbib]{revtex4-2}

\usepackage[utf8]{inputenc}
\usepackage{amsmath}
\usepackage{amssymb}
\usepackage{amsthm}
\usepackage{thmtools}
\usepackage{refcount}
\usepackage{siunitx}
\usepackage{graphicx}
\usepackage{hyperref}
\hypersetup{colorlinks,allcolors=black}
\usepackage{xcolor}

\usepackage{natbib}
 \usepackage[capitalise]{cleveref}
\usepackage{microtype}
\usepackage{bm}
\usepackage{lipsum}
\usepackage[english]{babel}
\usepackage{color}
\usepackage{graphicx}
\usepackage{xcolor}
\usepackage{braket}

\declaretheoremstyle[
headpunct={},
spaceabove=6pt, spacebelow=6pt,
headfont=\normalfont\bfseries,
notefont=\mdseries, notebraces={(}{)},
bodyfont=\normalfont,
postheadspace=1em,
]{exstyle}
\declaretheoremstyle[
spaceabove=6pt, spacebelow=6pt,
headfont=\normalfont\bfseries,
notefont=\mdseries, notebraces={(}{)},
bodyfont=\normalfont,
postheadspace=1em,
headpunct={},
qed=$\blacksquare$,
numbered=no
]{solstyle}

\graphicspath{{figs/}}

\begin{document}
\author{Rafi Mir-Ali Hessami}
\email{rafimah@stanford.edu}
\affiliation{Stanford PULSE Institute, SLAC National Accelerator Laboratory, Menlo Park, California 94025, USA}

\author{Haidar Al-Naseri}
\email{hnaseri@stanford.edu}
\affiliation{Stanford PULSE Institute, SLAC National Accelerator Laboratory, Menlo Park, California 94025, USA}

\author{Monika Yadav}
\affiliation{Department of Physics and Astronomy, University of California, Los Angeles, California 90095, USA}
\affiliation{Old Dominion University, Norfolk, Virginia 23529, USA}

\author{M. H. Oruganti}
\affiliation{Department of Physics and Astronomy, University of California, Los Angeles, California 90095, USA}

\author{B. Naranjo}
\affiliation{Department of Physics and Astronomy, University of California, Los Angeles, California 90095, USA}

\author{J. Rosenzweig}
\affiliation{Department of Physics and Astronomy, University of California, Los Angeles, California 90095, USA}

\title{Simulating Unruh Radiation in High--Intensity Laser--Electron Interactions for Near--Term Experimental Tests}

\begin{abstract}
The Unruh effect predicts that a uniformly accelerating observer perceives the vacuum as a thermal bath, yet direct observation remains elusive \cite{Unruh}. We simulate Unruh radiation in realistic high-intensity laser--electron collisions relevant to FACET-II and LUXE using fully three-dimensional Monte Carlo methods. In our model, Unruh emission is treated as scattering from a rest-frame thermal spectrum with Klein--Nishina cross sections, while nonlinear Compton radiation is computed across many harmonic orders with photon recoil. We map the laboratory-frame spectral--angular distributions and identify phase-space regions where the Unruh-to-Compton ratio is maximized. For current FACET-II-like parameters ($a_{0} = 5$), favorable windows for observing Unruh radiation occur at \mbox{200--400~$\mu\mathrm{rad}$} and \mbox{2--3~GeV}, although the absolute signal is small. For future LUXE Phase-1 ($a_{0} = 23.6$), the ratio increases by more than two orders of magnitude, with optimal angles around \mbox{800~$\mu\mathrm{rad}$} and photon energies \mbox{2--6~GeV}. Our results suggest that targeted off-axis, mid-energy selections can enhance sensitivity to Unruh-like signatures, motivating dedicated measurements and further theoretical scrutiny of the emission model at high field strengths.
\end{abstract}

\maketitle

\section{Introduction}

The Unruh effect predicts that a uniformly accelerating observer in a vacuum perceives the Minkowski vacuum as a thermal bath of particles with a temperature proportional to the observer's proper acceleration \cite{Unruh}. This phenomenon, derived from quantum field theory (QFT) in curved spacetime, remains experimentally unverified due to the extreme accelerations required to produce observable effects. Detecting Unruh radiation in a laboratory setting would constitute a significant test of fundamental quantum principles and provide insight into the interplay between acceleration, vacuum fluctuations, and particle production.

The Unruh effect has attracted growing interest in recent years, leading to a variety of proposed detection schemes. It has been suggested that the effect could be observed using storage rings \cite{StorageRing}, while other approaches have considered media with time-varying refractive indices \cite{RefractiveIndex}. Additional proposals include using Penning traps as potential detection platforms \cite{PenningTrap}, as well as employing electromagnetic waveguides to simulate phenomena from QFT in curved spacetime \cite{WaveGuide}.
Several detection concepts also build upon advances in high-intensity laser technology. Experimental proposals often focus on extreme field environments where electrons undergo rapid acceleration over short timescales. Among the most promising approaches is the interaction of relativistic electron beams with intense laser pulses\cite{chen1999testing, schutzhold2006signatures, Tajima2006}. In such scenarios, electrons may experience quantum transitions influenced by the Unruh effect.
With significant progress in laser intensity over recent decades, interest in these acceleration-based setups has continued to grow \cite{E-144, E320prep, Rutherford, laserfield}. Nevertheless,  distinguishing the Unruh signal from conventional radiation processes, such as nonlinear Compton scattering, remains a central challenge.

Previous studies have modeled Unruh radiation from a single particle in an idealized trajectory, often neglecting background radiation or beam dynamics \cite{Rotating_Unruh3,Rotating_Unruh,brodin2008laboratory}. In this work, we extend these approaches by incorporating realistic experimental geometries and performing fully three-dimensional Monte Carlo simulations of both Unruh and Compton photon production. Our modeling includes beam distributions matched to parameters from the existing facility FACET-II, enabling a direct assessment of experimental feasibility. We also examine parameters that we expect to be achieved at near-term facilities such as LUXE \cite{LUXETDR}.

Furthermore, we examine Unruh radiation in the laboratory frame, identifying specific regions in energy-angle phase space where Unruh photons can be better distinguished from background processes. These regions are characterized by a distinct emission pattern due to the Lorentz boost of thermal-like radiation, offering a potential experimental signature.

Although collective effects are neglected in the present simulations due to the low-density regime under consideration, we provide a theoretical analysis of possible collective enhancements in high-density beams using a Vlasov-based framework. This analysis lays the foundation for future studies of Unruh radiation in dense plasma environments.

This paper is organized as follows: In Sec.~\ref{sec:theory}, we review the theoretical background and calculate the Unruh phase space distribution in the lab frame, as well as the particle production rate. Section ~\ref{sec:exp_consideration} presents the simulation methods and beam-laser configurations used in this work. In Sec.~\ref{sec:conclusion}, we outline the implications for near-term experiments and future directions.

\section{Theory} \label{sec:theory}

To observe a measurable Unruh signal, the electron must experience external acceleration. In our scheme, this force arises from the field of a high-intensity laser. We consider a circularly polarized optical field throughout this work. Although the original derivation of Unruh radiation\cite{Unruh} assumes uniform linear acceleration, subsequent studies\cite{Rotating_Unruh,Rotating_Unruh2,Rotating_Unruh3} have shown that the instantaneous Unruh temperature $T_U$, defined as

\begin{equation} \label{eqn:TU}
    k_B T_U = \frac{\hbar a}{2 \pi c}
\end{equation}

where a is the magnitude of acceleration, provides a valid approximation even in non-uniformly accelerating frames. Moreover, because the laser field varies slowly compared to the photon formation timescale, we treat the electron’s acceleration as effectively constant during each interaction.

Classical radiation mechanisms, such as Larmor or synchrotron radiation, produce far more photons than Unruh radiation for any experimentally accessible case \cite{brodin2008laboratory, chen1999testing}, which can obscure the Unruh signal. However, these processes differ in their characteristic spectral-angular distributions, allowing for potential spectral and directional discrimination.
Our analysis begins in the electron rest frame, where the vacuum appears as a homogeneous, isotropic thermal photon bath at temperature $T_U$. We follow previous work\cite{brodin2008laboratory, mcdonald1998_unruh_larmor, schutzhold2010_quantum_radiation_lasers, rybicki1979radiative} in modeling the Unruh photon field as a uniform blackbody gas.

Scattering is modeled using the Klein–Nishina differential cross section\cite{weinberg1995qtf1}. By using the full Klein–Nishina formalism rather than the Thomson approximation, we relax the constraint $E_U \ll m_e c^2$ and instead adopt the weaker condition $E_U < m_e c^2$. This enables us to probe higher Unruh temperatures than in earlier works that assumed low-energy scattering\cite{brodin2008laboratory, yano2017_space_time_distortion, mcdonald1998_unruh_larmor, schutzhold2010_quantum_radiation_lasers}. Nevertheless, because the Unruh distribution is thermal and unbound, it is possible to sample a photon with energy exceeding the electron's total energy. We reject such unphysical events, which would correspond to a photon with more energy than the electron that emitted it.

This high-energy cutoff is physically motivated, as the energy of an emitted Unruh photon must be bound by the total energy of the emitting electron. Although the thermal spectrum formally includes arbitrarily energetic photons, a real electron can only emit photons up to its available lab frame energy, $\gamma m_e c^2$, where $\gamma$ is the Lorentz factor. We therefore enforce this as an upper bound. This ensures energy conservation and imposes kinematic consistency on the sampled spectrum. A more complete theoretical understanding of the Unruh effect at high temperatures where $T_U$ approaches or exceeds $m_e c^2$ remains an open question.

The rate of Unruh photon production in the lab frame is given by
\begin{equation} \label{eqn:numUnruhgamma}
    \frac{dN}{dt} = \frac{1}{\gamma} \sigma_{KN}^{averaged}\left(T_U\right) c\, n(T_U),
\end{equation}
where $\sigma_{KN}$ is the Klein–Nishina cross section, and $n(T_U)$ is the number density of Unruh photons in the rest frame. Assuming a thermal distribution, the spectral number density is given by
\begin{equation} \label{eqn:Unruh_blackbody}
    n_{E_U}(E_U, T_U) = \frac{1}{\pi^2 c^3} \frac{E_U^2}{e^{E_U / (k_B T_U)} - 1},
\end{equation}
and integrating over energy yields the total number density
\begin{equation} \label{eqn:Unruh_N}
    n(T_U) = \frac{2 \zeta(3) (k_B T_U)^3}{\pi^2 (\hbar c)^3},
\end{equation}
where $\zeta(3) \approx 1.202$ is the Riemann zeta function \cite{rybicki1979radiative}.

Because the Klein–Nishina cross section depends on photon energy, we compute an effective cross section by averaging over the thermal spectrum. Defining the dimensionless variable
\begin{equation} \label{eqn:R_def}
    R = \frac{E_U}{m_e c^2},
\end{equation}
the Klein–Nishina cross section is given by
\begin{multline} \label{eqn:sigma_KN}
    \sigma_{\mathrm{KN}}\left(E_U\right) =
    2 \pi r_0^{2} 
    \Bigg[
      \frac{1 + R}{R^{3}}
      \left(
        \frac{2 R (1 + R)}{1 + 2 R} 
        - \ln (1 + 2 R)
      \right) \\
      + \frac{\ln (1 + 2 R)}{2 R} 
      - \frac{1 + 3 R}{(1 + 2 R)^{2}}
    \Bigg],
\end{multline}
where $r_0$ is the classical electron radius. The thermally averaged cross section is
\begin{equation} \label{eqn:sigma_KN_avg}
    \sigma_{\mathrm{KN}}^{\mathrm{averaged}}\left(T_U\right) =
    \frac{
        \displaystyle \int n_{E_U}(E_U, T_U) \, \sigma_{\mathrm{KN}}(E_U) \, dE_U
    }{
        \displaystyle \int n_{E_U}(E_U,T_U) \, dE_U
    }.
\end{equation}
This weighted average accounts for the full range of photon energies in the Unruh bath and provides a consistent estimate for the scattering probability.

Once an emission event is triggered by this rate, we sample the energy and direction of the emitted Unruh photon from the spectral-angular distribution. In the rest frame, this distribution is given by
\begin{equation} \label{eqn:Planckspec}
    f(E_U, \xi, \phi) = \frac{1}{4\pi} \frac{E_U^2}{\hbar^3 \pi^2 c^3} \frac{1}{e^{E_U / (k_B T_U)} - 1},
\end{equation}
where $E_U$ is the photon energy, $\xi$ is the polar angle relative to the $z$-axis, and $\phi$ is the azimuthal angle. This expression gives the number of photons per unit volume, per unit energy, per unit solid angle.

To obtain the corresponding lab-frame distribution, we apply the Lorentz transformation detailed in Appendix~\ref{Lorentz}
\begin{equation} \label{eqn:plancktransform}
    f'(E_U', \xi', \phi') = \frac{f(E_U, \xi, \phi)}{\gamma - \xi' \sqrt{\gamma^2 - 1}},
\end{equation}
where primed quantities are defined in the lab frame. This transformation yields the angular-spectral distribution of Unruh photons as seen by the lab observer. The sampled photon’s energy and momentum are subtracted from the electron that emitted it, ensuring conservation of energy and momentum.

\section{Experimental considerations} 
\label{sec:exp_consideration}

To assess the feasibility of detecting Unruh radiation in planned experiments, we simulated photon production using realistic beam and laser parameters. This section outlines our numerical approach, including modeling of Unruh and Compton processes, and presents results for the FACET-II and LUXE facilities.

\subsection{Numerical methods}

To realistically simulate Unruh radiation in current experimental regimes, we developed a laser–electron beam interaction model capable of producing both Unruh and nonlinear Compton radiation. The simulation was implemented in MATLAB \cite{MATLAB}, and neglects collective effects as discussed in \cref{CollectiveEffectsSection}, since the electron densities under consideration remain well below the critical threshold of $10^{21}~\mathrm{cm}^{-3}$ at which the back-reaction of the electron dynamics on the external field start to be important. Consequently, the model described in \cref{sec:theory} is adopted, which is used to model Unruh radiation alongside Compton radiation.

The simulation propagates an electron beam in 3D space, formatted according to the \textsc{Lucretia} convention\cite{Lucretia}, through a focused laser field. The laser field was modeled using a simplified analytic approximation based on the Richards–Wolf formalism, modeling transverse and longitudinal components of the focused beam \cite{RichardsWolf1959}. For each macroparticle, the local field amplitude is used to compute the normalized vector potential $a_0$, given by

\begin{equation} \label{eqn:a0}
    a_0 = \frac{eE}{mc\omega},
\end{equation}

\noindent where $E$ is the local electric field amplitude, $m$ is the electron mass, and $\omega$ is the laser angular frequency\cite{E320prep}.

With both $a_0$ and the Lorentz factor $\gamma$ of each macroparticle known, the rates of both Compton and Unruh radiation can be computed. The total cross section of nonlinear Compton scattering is computed by calculating and summing the cross sections of all possible Compton scattering events up to a threshold minimum cross section value, after which the contribution is considered trivial. An event is determined by using this total cross section. Then, the specific order of the process is determined by using $\sigma\left(n\right)$ as a distribution function to sample from. To ensure that higher order processes, particularly rare high energy photon emissions, are accurately sampled, we employ a log-biased importance sampling algorithm. This technique preferentially samples the photon energy fraction $x = \frac{E_{\gamma}}{E_e}$, where $E_{\gamma}$ is the outgoing photon energy in the lab frame and $E_e$ is the initial electron energy in the lab frame, on a logarithmic scale to improve the efficiency of Monte Carlo integration across several orders of magnitude. However, this biasing requires a correction factor in the form of a Jacobian weight and can introduce small statistical errors in the total photon yield if undersampling occurs in regions with lower emission probability. To control this, we impose a cap on the acceptable discrepancy in the integrated photon number, limiting it to no more than 2\% relative error in all simulations \cite{Rubinstein2011}.

Following the treatment in Ref.~\cite{blackburn2015a}, the differential probability rate for nonlinear Compton scattering is calculated. This is expressed as

%\begin{multline} \label{eqn:probrate_Blackburn}
%    \frac{\partial W_n}{\partial x} = \frac{2\alpha m^2}{\pi %\varepsilon} \Biggl( \int^{\pi/2}_0 \Bigl[ -A^2_0 
%    + a^2_0\left(1 - x + \frac{1}{1 - x} \right) \\
%    \times \left(A^2_1 - A_0 A_2 \right) \Bigr] \, d\theta \Biggr)
%\end{multline}

%\noindent for linear polarization and

\begin{multline} \label{eqn:probrate_Blackburn}
    \frac{\partial W_n}{\partial x} = \frac{\alpha m^2}{4 \varepsilon} \Biggl( \Bigl[ -4J_n^2 
    + a_0^2\left(1 - x + \frac{1}{1 - x} \right) \\
    \times \left(J^2_{n-1} - J^2_{n+1} - 2J^2_n \right) \Bigr]\Biggr)
\end{multline}

for circular polarization, where $\frac{\partial W_n}{\partial x}$ is the differential probability with respect to outgoing photon energy and $n$ is the harmonic order. For further details on the argument of $J_n$ in Eq. \ref{eqn:probrate_Blackburn}, please see Ch. 2 in Ref.~\cite{blackburn2015a}.  

% Should I define A_i in the paper? Or can I just reference the source?

%$A_i$ is defined by

%\begin{equation} \label(eqn:Ai_def)
%    A_i \left(n,a,b\right) = \frac{1}{\pi} \int^{\pi}_0 \cos^i\left(\phi\right) \cos \left(\left(a + 2b\cos\left(\phi\right) \right)\sin\left(\phi\right) - n\phi \right) \partial \phi
%\end{equation}

%where a and b are given in equations \ref{eqn:adef} and \ref{eqn:bdef}, 

For each macroparticle, the probability of Compton scattering is evaluated across all relevant orders and if scattering is determined to occur, a corresponding macrophoton is generated with properties sampled according to the angular and energy distributions in Ref.~\cite{blackburn2015a}, and reproduced below

\begin{equation} \label{eqn:angledist_blackburn}
    \hbar \omega' = \frac{nu}{1 + nu - \left( 1 + nu - \frac{4\gamma^2}{1+a_0^2} \right)\sin^2{\frac{\theta}{2}}}\varepsilon,
\end{equation}

\begin{equation} \label{eqn:u_blackburn}
    u = \frac{2k \cdot q}{q^2} \approx \frac{4\varepsilon \hbar \omega'}{m^2 \left(1 + a_0^2 \right)},
\end{equation}

\noindent where $k$ and $q$ are the four-momenta of the laser photon and the electron, respectively, and $\theta$ is the photon scattering angle. Importantly, the photon emission direction is not assumed to be collinear with the electron velocity. Photon recoil is applied to the electron, leading to a broadened, low-energy tail in the electron spectrum.

Unruh radiation is modeled using a similar Monte Carlo framework. The emission probability is given by Eq.~\ref{eqn:numUnruhgamma}, and the spectral–angular properties of emitted Unruh photons are computed using Eq. \ref{eqn:plancktransform}.

\subsection{Simulation results: FACET-II parameters}

In this subsection, we consider the Unruh radiation in the electron-laser collision using the parameters in FACET-II.  
At $a_0 = 5$, nonlinear Compton scattering dominates the total photon yield, producing a sharply forward-peaked spectrum. The Unruh photon spectrum, though much weaker in absolute rate, is comparatively broader in both energy and angle. The parameters of the electron beam and laser used for these simulations are given in Table~\ref{tab:collision_params}.

\begin{table}[t]
  \caption{Simulation parameters for the electron–laser collision at FACET-II.}
  \label{tab:collision_params}
  \centering
  \begin{tabular}{ll}
    \hline
    Parameter & Value \\
    \hline
    Electron beam energy & 10.0~GeV \\
    Electron beam energy spread (rms) & 0.182\% \\
    Electron beam charge & 2.00~nC \\
    Electron beam spot size in $x$ (rms) & 24.3~$\mu\mathrm{m}$ \\
    Electron beam spot size in $y$ (rms) & 29.6~$\mu\mathrm{m}$ \\
    Electron beam length (rms) & 78.8~$\mu\mathrm{m}$ \\
    Laser $a_{0}$ & 5.00 \\
    Laser f-number & 2.00 \\
    Laser pulse length & 45.0~fs \\
    Collision angle & 30.0~degrees \\
    Peak Unruh temperature & 0.113~MeV \\
    \hline
  \end{tabular}
\end{table}

\begin{figure}[h]
    \centering
    \includegraphics[width=\linewidth]{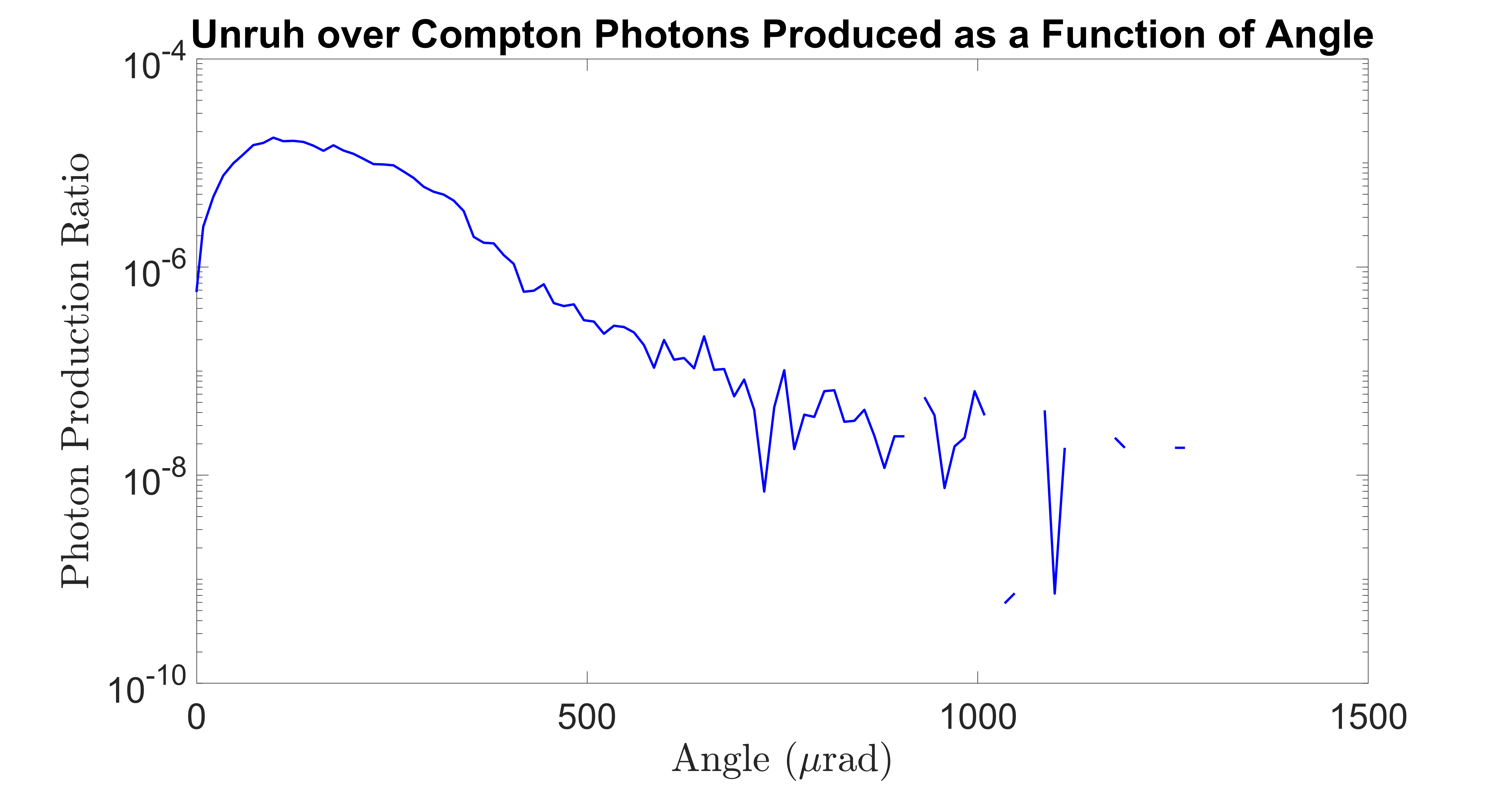}
    \caption{Angular dependence of the Unruh-to-Compton photon density ratio at $a_0 = 5$.}
    \label{fig:FACET-II_ratio}
\end{figure}
While the Compton signal dominates absolutely, important structure emerges when examining the ratio of Unruh to Compton photon density. As shown in Figure ~\ref{fig:FACET-II_ratio}, the angular dependence of this ratio features a broad peak near $150\,\mu\mathrm{rad}$, reaching values around $10^{-5}$. This region represents a relative minimum in Compton yield while the Unruh distribution remains relatively high. At angles greater than $\sim400\,\mu\mathrm{rad}$, the ratio falls off due to diminishing photon yield in both channels. Outside of this window, especially at very small angles ($<50\,\mu\mathrm{rad}$), the Compton spectrum overwhelms the signal, while at large angles ($>700\,\mu\mathrm{rad}$), both signals become too weak to be viable for detection.

\begin{figure}[h]
    \centering
    \includegraphics[width=\linewidth]{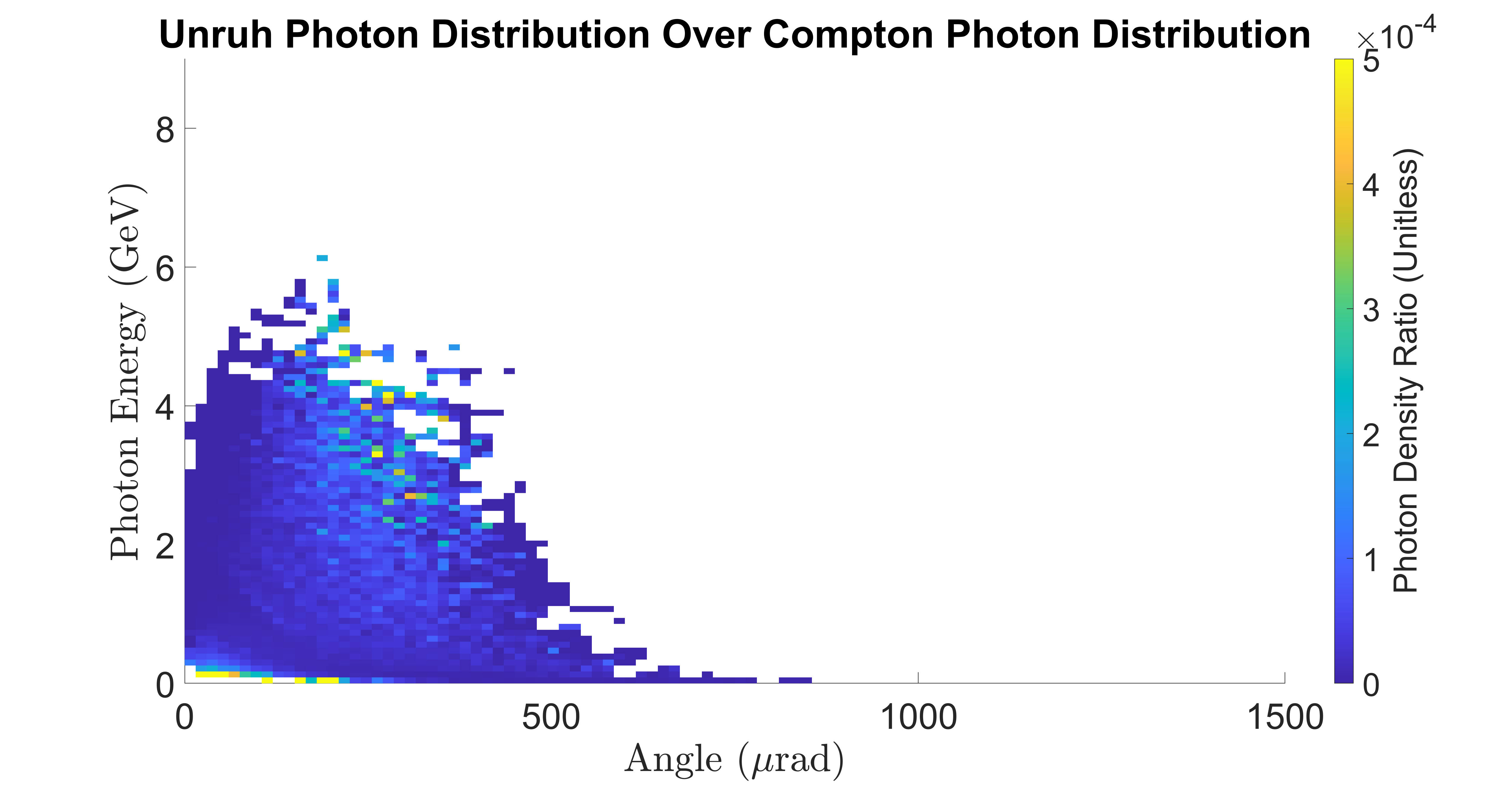}
    \caption{2D distribution of Unruh-to-Compton photon density ratio as a function of energy and angle at $a_0 = 5$.}
    \label{fig:FACET-II_ratio_2D}
\end{figure}

The corresponding two-dimensional energy-angle ratio map (Figure ~\ref{fig:FACET-II_ratio_2D}) shows that the most favorable detection window lies in the higher energy range (2–3\,GeV) and moderate angles (200–400\,$\mu\mathrm{rad}$). In this band, the Unruh signal grows more quickly than Compton. This suggests that, despite the small absolute signal, careful filtering in both angle and energy can reveal regions with locally enhanced Unruh signal-to-background ratios. This strategy is especially important at modest $a_0$, where the signal is not intrinsically strong, but becomes potentially visible due to background suppression and high detector sensitivity.

To evaluate whether higher-intensity lasers improve observability, we next simulate the LUXE experiment, which operates at a significantly larger $a_0$.

\subsection{Simulation results: LUXE parameters}
In this subsection, we consider the Unruh signal in the collision of 16.5 GeV electron beam and 350 TW laser beam that is planned to take place at the LUXE experiment at DESY in phase-1. The parameters of the electron beam and laser used for these simulations are given in Table~\ref{tab:collision_params_LUXE}.

\begin{figure}[h]
    \centering
    \includegraphics[width=\linewidth]{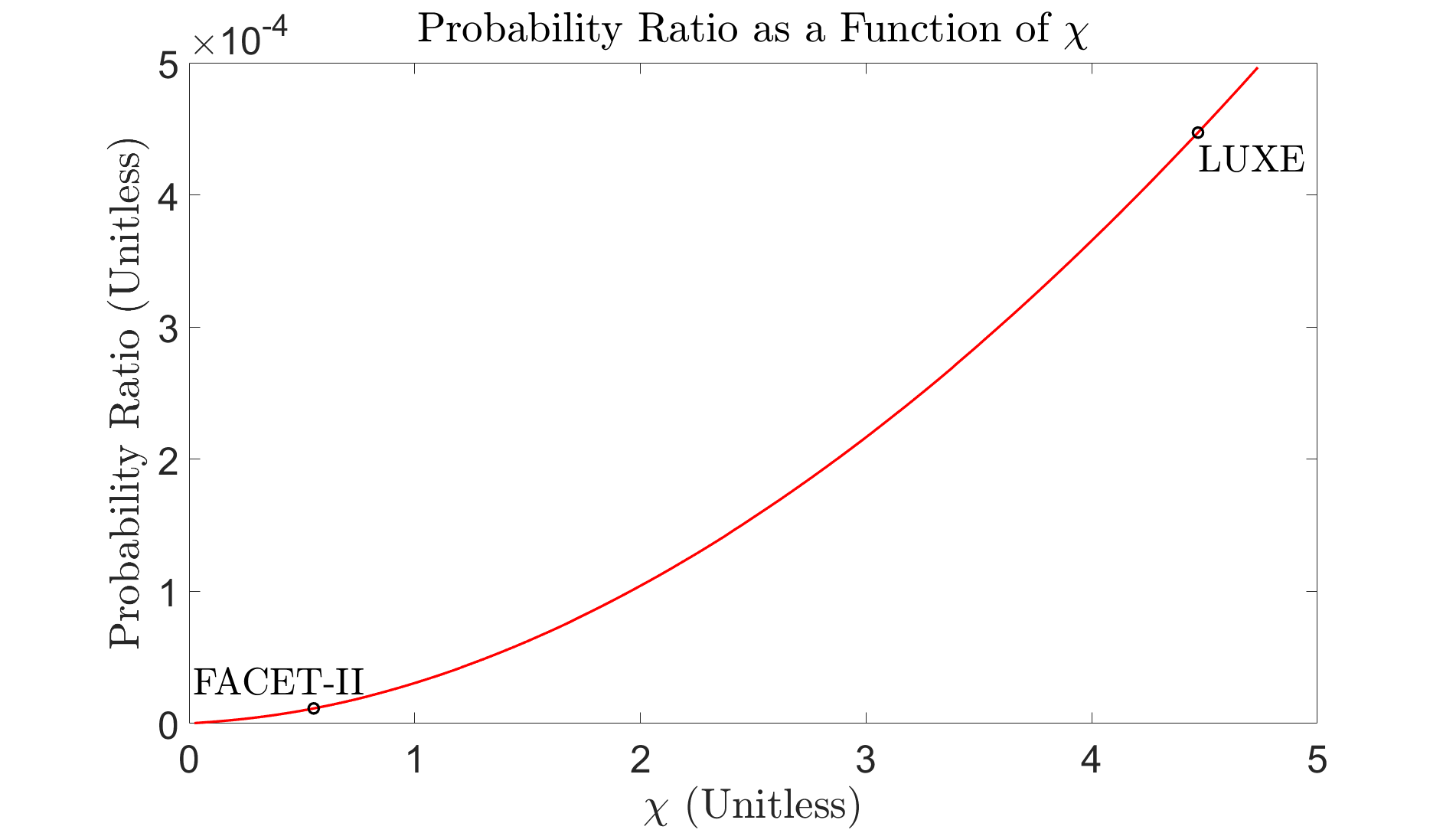}
    \caption{Ratio of Unruh to Compton differential photon production rates $\mathrm{d}N/\mathrm{d}t$ as a function of the nonlinear parameter $\chi$.}
    \label{fig:prob_rate_vs_chi}
\end{figure}

With the higher normalized vector potential of $a_0 = 23.6$, the Compton spectrum becomes even more intense, with significantly increased high-energy photon yield. However, the Unruh-to-Compton ratio improves dramatically because the differential production rate for Unruh photons grows more rapidly with the quantum parameter $\chi$ than the differential production rate for Compton photons.
This scaling is illustrated in Figure ~\ref{fig:prob_rate_vs_chi}, which shows the Unruh-to-Compton production rate ratio
\begin{equation}
\frac{(\mathrm{d}N/\mathrm{d}t)_{\mathrm{Unruh}}}{(\mathrm{d}N/\mathrm{d}t)_{\mathrm{Compton}}}
\end{equation}
as a function of $\chi$. The curve rises steeply, reflecting the favorable scaling of Unruh-like radiation processes relative to nonlinear Compton scattering as field strength increases.

\begin{figure}[h]
    \centering
    \includegraphics[width=\linewidth]{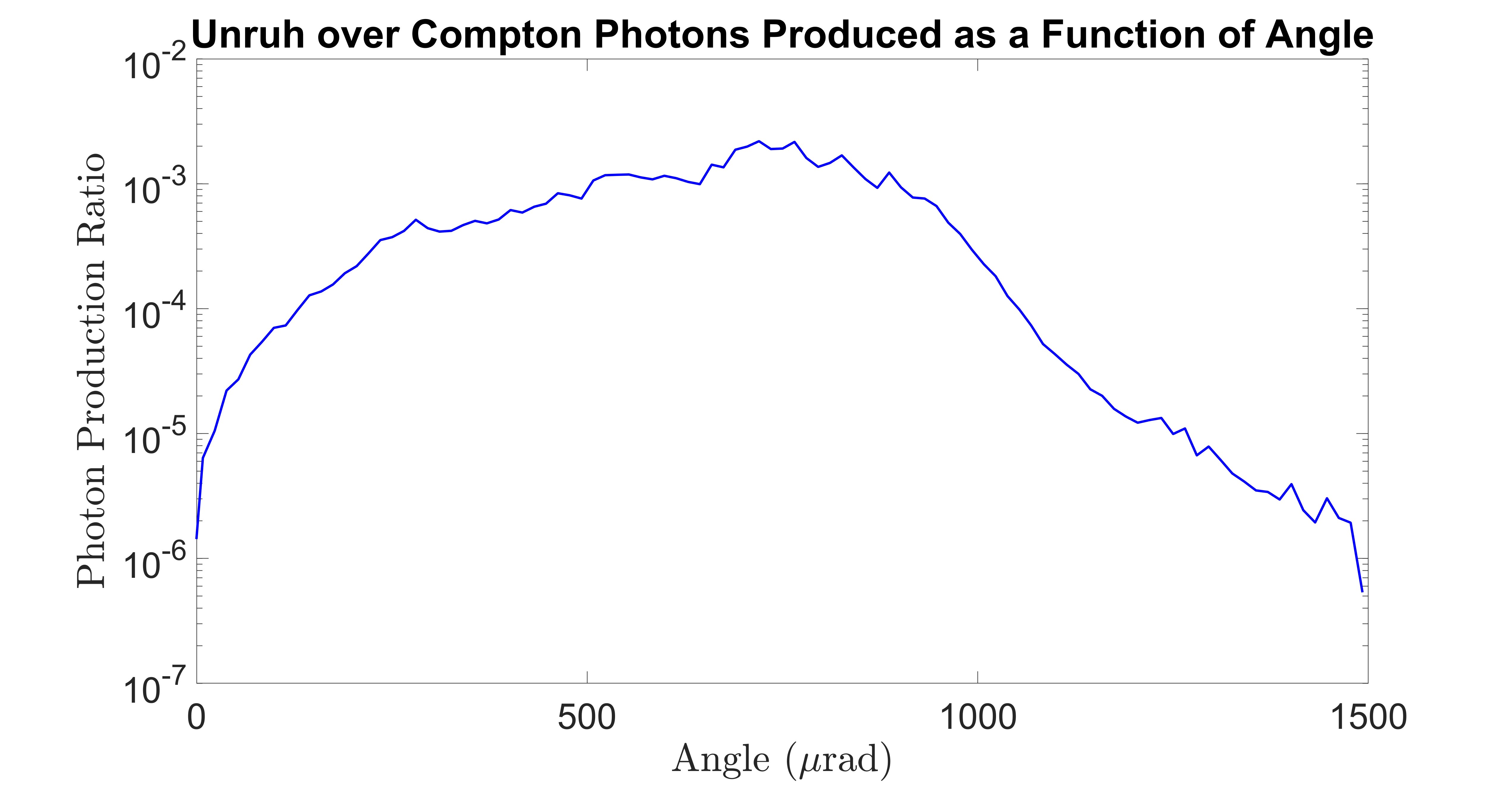}
    \caption{Angular dependence of the Unruh-to-Compton photon density ratio at $a_0 = 23.6$.}
    \label{fig:LUXE_ratio}
\end{figure}

The impact of this scaling is visible in the angle-only plot of the photon density ratio (Figure ~\ref{fig:LUXE_ratio}). The peak shifts to slightly higher angles, around $800\,\mu\mathrm{rad}$, and reaches values exceeding $10^{-3}$, a significant improvement over the FACET-II case. The slope beyond the peak is gentler, indicating a broader region of enhanced detectability extending around $800\,\mu\mathrm{rad}$.

\begin{figure}[h]
    \centering
    \includegraphics[width=\linewidth]{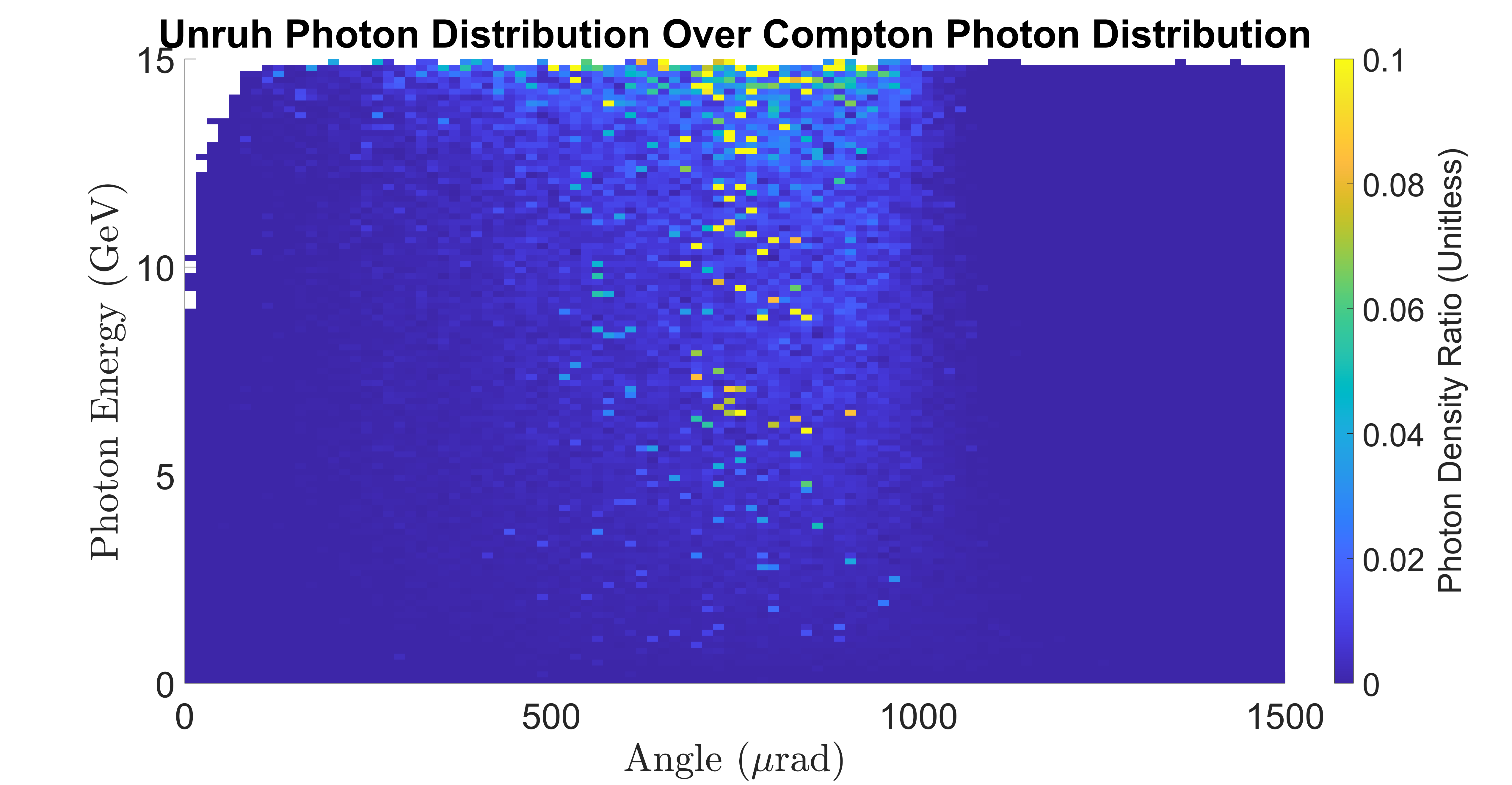}
    \caption{2D distribution of Unruh-to-Compton photon density ratio as a function of energy and angle at $a_0 = 23.6$.}
    \label{fig:LUXE_ratio_2D}
\end{figure}

The 2D ratio map (Figure ~\ref{fig:LUXE_ratio_2D}) confirms that the high-intensity regime opens new windows of observability, now including higher energy photons in the 5–10\,GeV range and angular offsets centered around 800\,$\mu\mathrm{rad}$. The spatial and spectral phase space of Unruh visibility expands, making detection increasingly plausible as $a_0$ rises. It is worth noting that at very high energies close to the energy of the beam, this model cannot accurately capture the physics, as that corresponds to Unruh temperatures very close to the electron rest mass. 

Overall, the simulations at both $a_0 = 5$ and $a_0 = 23.6$ indicate that experimental efforts should focus on moderate angles ($200\text{–}400\,\mu\mathrm{rad}$ at low intensity and up to $800\,\mu\mathrm{rad}$ at high intensity) and photon energies below 6 GeV. These regions maximize the ratio of Unruh to Compton radiation and define practical detection bands where background suppression techniques, such as angular collimation and energy filtering, may be most effective.

\begin{table}[t]
  \caption{Simulation parameters for the electron–laser collision at LUXE phase~1.}
  \label{tab:collision_params_LUXE}
  \centering
  \begin{tabular}{ll}
    \hline
    Parameter & Value \\
    \hline
    Electron beam energy & 16.5~GeV \\
    Electron beam energy spread (rms) & 0.100\% \\
    Electron beam charge & 0.250~nC \\
    Electron beam spot size in $x$ (rms) & 5.00~$\mu\mathrm{m}$ \\
    Electron beam spot size in $y$ (rms) & 5.00~$\mu\mathrm{m}$ \\
    Electron beam length (rms) & 5.00~$\mu\mathrm{m}$ \\
    Laser $a_{0}$ & 23.6 \\
    Laser f-number & 3.75 \\
    Laser pulse length & 25.0~fs \\
    Collision angle & 20.0~degrees \\
    Peak Unruh temperature & 0.534~MeV \\
    \hline
  \end{tabular}
\end{table}

\section{Conclusion} \label{sec:conclusion}

We have presented fully three-dimensional Monte Carlo simulations of Unruh and nonlinear Compton radiation in realistic electron-laser collisions, using parameters relevant to current and near-future experimental capabilities. By comparing results at two different experimental facilities, FACET-II and LUXE,  we assessed the detectability of Unruh radiation in the laboratory frame and identified specific regions of phase space where its signal is relatively enhanced.

For FACET-II, with $a_0 = 5$, the Unruh photon yield remains deeply buried beneath the Compton background across the full angular and spectral range. Even in the most favorable regions, moderate angles around $200$–$400\,\mu\mathrm{rad}$ and photon energies below 3\,GeV, the signal-to-background ratio remains extremely low, making detection at this intensity unlikely with existing instrumentation.

For the LUXE experiment, we used $a_0 = 23.6$. Here, the situation improves significantly. The Unruh-to-Compton photon production rate grows more rapidly with increasing $\chi$, leading to a broader and more pronounced region of favorable contrast. In particular, angular offsets of $200$–$800\,\mu\mathrm{rad}$ and photon energies between 2–6\,GeV offer promising detection windows, where the relative Unruh signal exceeds $ 10^{-3}$. While still challenging, detection in this regime may be within reach of carefully designed experiments.

We therefore suggest that upcoming experiments operating in similar high-intensity parameter spaces explicitly target these off-axis, mid-energy phase space regions. Doing so may enable the first observation of Unruh radiation and provide an experimental probe of quantum vacuum effects under acceleration, constituting an important step forward in connecting theoretical predictions with measurable signatures.

\section{Acknowledgements}

The authors gratefully acknowledge David Reis, Agostino Marinelli, and Alexander Knetsch for insightful discussions and feedback that contributed significantly to the development of the ideas and analysis presented in this work.

\appendix

\section{Collective effects} \label{CollectiveEffectsSection}
In this section, our focus is on investigating how the collective effects of the electron beam affects the number of Unruh photons scattered back in the lab frame. In the main text of this work, we used parameters of electron beams that are not dense enough to justify the inclusion of collective effects. However, there are existing and planned facilities where the density of the electron beam is high enough to make collective effects relevant. Furthermore, no previous work has included collective effects in the study of Unruh radiation in laser-plasma interaction. Thus, it is worth paying attention to how the collective effects of the electron beam might affect the Unruh temperature and hence the number of scattered photons.

In our study of the collective effects, we use the relativistic Vlasov equation as a kinetic equation. For the high-intensity laser, we consider a circularly polarized field
\begin{align}
    E_x&= E_0 \cos (\omega t) \mathbf{e}_x\\
    E_y&= E_0 \cos (\omega t) \mathbf{e}_y
\end{align}
where $E_0$ is the field amplitude and $\omega$ frequency of the laser.

Using this model of the laser field in the relativistic equation, we derive
\begin{equation}
\label{CircularPol}
    \frac{\partial f_e}{\partial t}
    +  qE_x \frac{\partial f_e}{\partial p_x}+  qE_y \frac{\partial f_e}{\partial p_y}=0
\end{equation}
This system is closed by Ampère's law
\begin{align}
\label{Ampers}
    \partial_tE_x&=-4\pi c^2 e \int d^3p \frac{p_x}{\epsilon}f_e\\
    \partial_tE_y&=-4\pi c^2 e \int d^3p \frac{p_y}{\epsilon}f_e
\end{align}
where $\epsilon=\sqrt{1+p^2}$. Note that the ions are considered to be a neutralizing background and we do not consider their motion. 

The main information that we are interested in acquiring from the solution of the relativistic Vlasov system is the phase-space averaged acceleration. The energy spectrum of the Unruh radiation is dependent on the Unruh temperature $T_U$, which in turn depends on the acceleration $a$. The phase-space averaged acceleration is 
\begin{equation}
    \mathbf{a}=\int d^3p\bigg( \frac{ \mathbf{F}}{\epsilon}-\frac{\mathbf{F}\cdot \mathbf{v}\mathbf{v}}{c^2\epsilon}\bigg)f_e(\mathbf{p},t)
\end{equation}
where the force is $\mathbf{F}=e\mathbf{E}$ and $\mathbf{v}=\frac{\mathbf{p}}{\epsilon}$. 

For the circularly polarized case considered in our work, we find
\begin{align}
\label{Acceleration}
    a_x&=\int d^3p \bigg(\frac{E_x}{\epsilon}-\frac{c^2p_x}{\epsilon^3} (E_xp_x+E_yp_y)\bigg)f_e\notag\\
    a_y&=\int d^3p \bigg(\frac{E_y}{\epsilon}-\frac{c^2p_y}{\epsilon^3} (E_xp_x+E_yp_y)\bigg)f_e
\end{align}
and the total acceleration is then given by $a=\sqrt{a_x^2+a_y^2}$.

We solve the kinetic Vlasov equation together with Ampère's law numerically. The numerical solution is simplified by using the canonical transformation

\begin{align*}
q_{x}&=p_{x} +eA_{x}\\
q_{y}&=p_{y} +eA_{y}
\end{align*}
Using the canonical transformation in \cref{CircularPol}, we get
\begin{equation}
\label{Vlasov_canon.}
    \frac{\partial f_e }{\partial t}(q_{x},q_{y},p_{z},t)=0
\end{equation} 

Given the initial values of the numerical setup, for any distribution function that is initially $f_e=F(q_x,q_y,q_z)$, the time-dependent solution to \cref{Vlasov_canon.} is $F(q_x-A_x(t),q_y-A_y(t),p_z)$. Here, $A_x$ and $A_y$ are the sum of external and self-consistent vector potential.

Since we are considering the interaction of plasma with external lasers, we can divide the total field into two parts, 

\begin{align}
A_{tot}&= A_{ex}+A_{s}\\
    E_{tot}&= E_{ex}+ E_s
\end{align}
where $E_{ex}$ is the external field that describes the laser pulse that is given in the system, which does not satisfy Ampère's law, and $E_s$ is the field that is determined by Ampère's law. For the electron beam, we can use any relativistic thermodynamic equilibrium distribution function in momentum space to represent the initial beam. Since we should solve the relativistic Vlasov equation in the rest frame of the electron, we should consider a Lorentz-boosted distribution function.

We allow the beam to have a boost in the momentum space $p_0$ in the $z$-direction, perpendicular to the laser polarization. Using the Synge-J{\"u}ttner distribution, the electron beam in the lab frame is given by 
\begin{equation}
    f_\text{beam}= \frac{e^{-\sqrt{1+q_{\bot}^2+ (p_z-p_0)^2}/E_{th}} }{\int d^3pe^{-\sqrt{1+q_{\bot}^2+ (p_z-p_0)^2}/E_{th}} }
\end{equation}
where $E_{th}=\sqrt{1+p_{th}^2}-1$. Here $p_{th}$ is thermal momentum and $q_{\bot}=\sqrt{q_x^2+q_y^2}$.

Then, we use the following Lorentz transformation
\begin{align*}
    q_x'&=q_x\\
    q_y'&=q_y\\
    p_z'&=\gamma_b\big(p_z-v\sqrt{1+p^2} \big)
\end{align*}
where $\gamma_b=1/\sqrt{1-v^2}$, here $v$ is the velocity of the electron beam. The electron beam in the rest frame is now 
\begin{equation}
\label{f_rest}
    f_\text{beam}= \frac{\exp\bigg(-p_0\big[\sqrt{1+q^{\prime 2}_{\bot} + p_z'}-1\big]/E_{th}\bigg) }{\int d^3p\, \exp\bigg(-p_0\big[\sqrt{1+q^{\prime 2}_{\bot}+ p_z'}-1\big]/E_{th}\bigg) }
\end{equation}
The distribution function is now more localized in the momentum space compared to the lab frame.

The solution to the Vlasov system is obtained by using \cref{f_rest} as an initial value of $f_e$ in \cref{Vlasov_canon.} and solving \cref{Ampers} for each time-step.

\begin{figure}
    \centering
    \includegraphics[width=\linewidth]{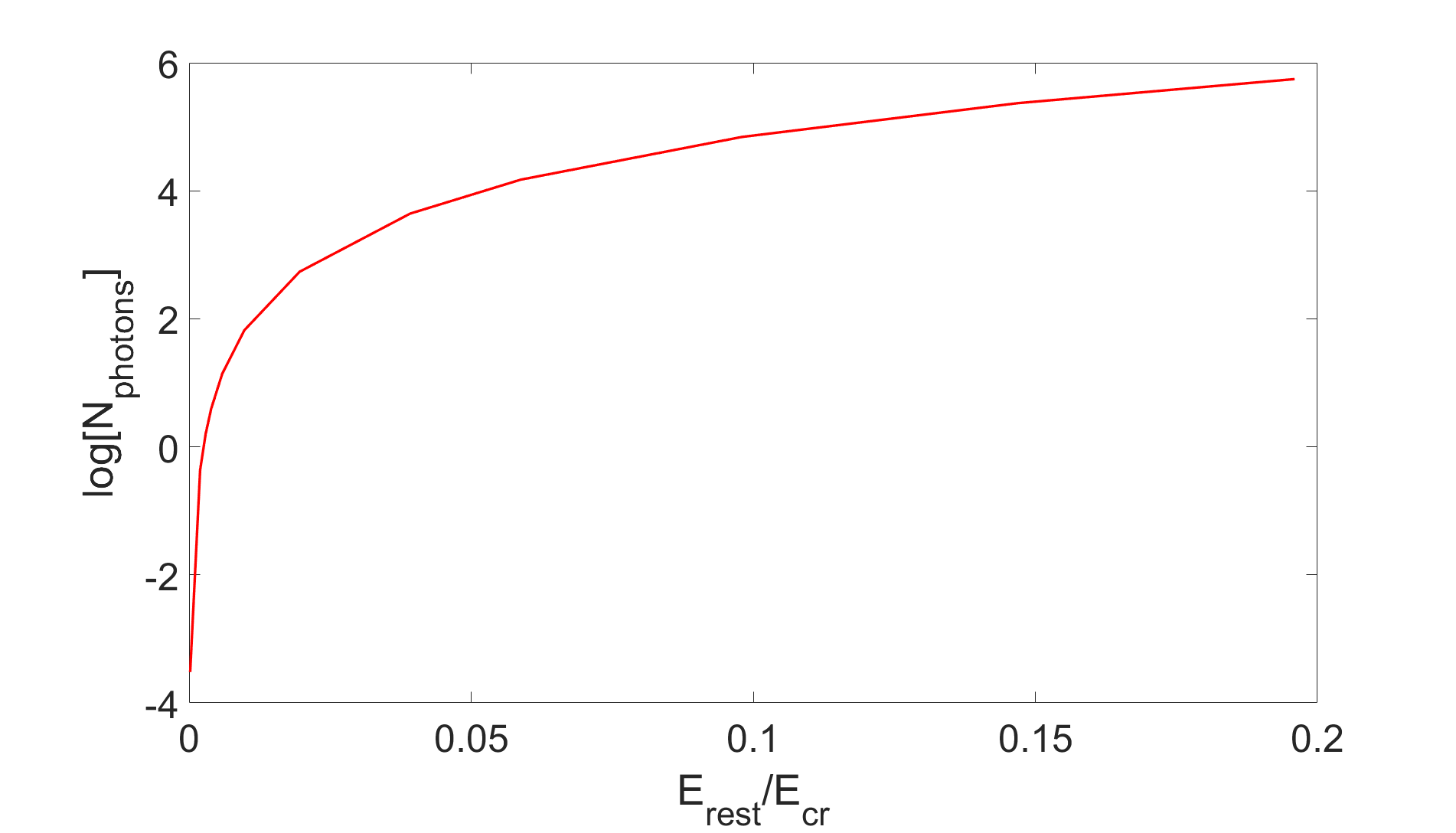}
    \caption{Number of scattered photons in the lab frame during one laser period as a function of the field amplitude in the rest frame. We have used $T_{plasma}$ = 510 MeV, $\lambda$ = 800 nm and a plasma density of $n_0=7\times 10^{18} \textrm{ cm}^{-3}$. }
    \label{Fig1}
\end{figure}

Now, we present the results of the collective effects, where we study how the number of scattered photons $N$ is dependent on the initial values of the numerical setup. In \cref{Fig1}, we plot the number of Unruh photons as a function of the electric field in the rest frame of the electron. So, what we have on the x-axis of \cref{Fig1} can be seen as the $\chi$ value that the electron experiences.

We used an electron beam with the density $n_0=7\times 10^{18}\textrm{ cm}^{-3}$, a $800$ nm laser wavelength, and a plasma temperature of $510$ MeV. As expected, stronger fields lead to stronger accelerations and therefore higher Unruh temperatures. In particular, we have the greatest increase in the number of photons $N$ when we go from $\chi=0.001$ to $\chi=0.01$. For higher $\chi>0.1$, we have a slower increase in the number of photons because electrons will approach the speed of light quite quickly. This would reduce the time that an electron is affected by the external force, and hence the increase in the number of photons is slower.

\begin{figure}
    \centering
    \includegraphics[width=\linewidth]{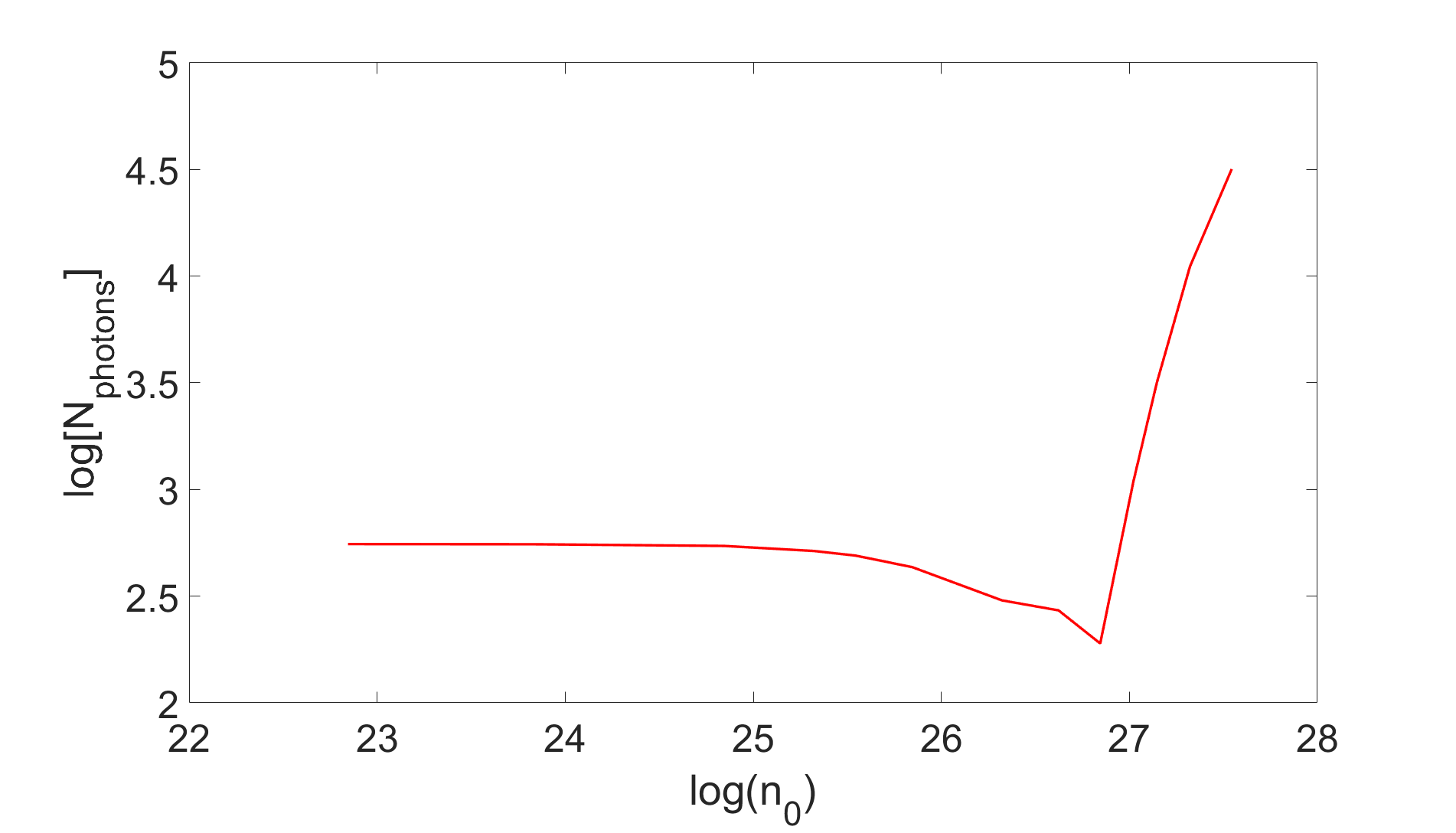}
    \caption{Number of scattered photons in the lab frame per electron during one laser period as a function of the plasma density. We have used an energy spread of 510 MeV,  $\lambda$ = 800 nm and the values of the electric field in the rest frame is $E_{rest}$ = 0.02.}
    \label{Fig2}
\end{figure}

To investigate how the plasma density affects the number of scattered Unruh photons, we plot the number of Unruh photons $N$ as a function of the plasma density in \cref{Fig2}. As the plasma becomes more dense, the back-reaction of the plasma on the laser field starts to become more important. The net acceleration of the electrons decreases once the back-reaction becomes significant. This is because the plasma starts to screen the laser field and the total field experienced by the electrons is lower than the input laser field. However, for a more dense plasma, the plasma frequency starts to increase significantly. In particular, for $n_0 \ge 10^{21} \textrm{ cm}^{-3}$, the plasma frequency is higher than the laser frequency. This will lead to a superposition of the laser and a self-consistent field, which results in a stronger total field. This results in a higher acceleration $a$ and, hence, a higher Unruh temperature.

\section{Lorentz-transformation} \label{Lorentz}

In this appendix, we present the derivation of the Lorentz transformation and how the spectral-angular distribution of the Unruh photon is transformed. 

We begin with equation \ref{eqn:Planckspec}. We then boost this into the lab frame in equation \ref{eqn:jacobian}.

\begin{equation} \label{eqn:jacobian}
    f'\left( E', \xi', \phi' \right) = f\left(E,\xi,\phi \right) \left| \frac{\partial \left(E,\xi,\phi \right) }{\partial \left( E', \xi', \phi' \right) } \right|
\end{equation}

To calculate the Jacobian, we calculate the transformation from the beam frame to the lab frame. 

First, we consider a photon of energy E in the direction of travel

\begin{equation}
    \hat{n} = \hat{x}\sqrt{1-\xi^2} \cos{\phi} + \hat{y} \sqrt{1 - \xi^2} \sin{\phi} + \hat{z} \xi
\end{equation}

\begin{equation} \label{eqn:lambda}
    k'^{\mu} = 
    \begin{pmatrix} \gamma & 0 & 0 & \sqrt{\gamma^2 - 1} \\
    0 & 1 & 0 & 0 \\
    0 & 0 & 1 & 0 \\
    \sqrt{\gamma^2 - 1} & 0 & 0 & \gamma \\
    \end{pmatrix}
    k^{\mu}
\end{equation}

Assuming $\gamma \gg 1$, we extract the transformations for the three variables.

\begin{equation}
    E' = E \gamma \left(1 + \xi \left(1 - \frac{1}{2\gamma^2} \right) \right)
\end{equation}

\begin{equation}
    \xi' = 1 - \frac{1}{2 \gamma^2} \left(\frac{1-\xi}{1+\xi} \right)
\end{equation}

\begin{equation} 
\phi' = \phi 
\end{equation}

where we used $\gamma \gg 1$.

Finally, after transforming the distribution, we calculate the Unruh spectrum per photon per electron from the lab-frame distribution function.

\bibliography{refs}

\end{document}